\documentclass[journal]{IEEEtran}
\ifCLASSINFOpdf
\else
\fi
\RequirePackage[latin1,utf8]{inputenc}
\usepackage[T1]{fontenc}
\usepackage{graphicx}
\usepackage{cite}
\usepackage{picinpar}
\usepackage{amsmath}
\usepackage{amssymb,bm}
\usepackage{float}
\usepackage{stfloats}
\usepackage{url}
\usepackage[latin1]{inputenc}
\usepackage{colortbl}
\usepackage{soul}
\usepackage{multirow}
\usepackage{pifont}
\usepackage{color}
\usepackage{alltt}
\usepackage{enumerate}
\usepackage{siunitx}
\usepackage{hyperref} 
\usepackage{epstopdf}
\usepackage{pbox}
\usepackage[linesnumbered,ruled]{algorithm2e}
\usepackage[font=footnotesize]{caption}
\usepackage[font=footnotesize]{subcaption}
\usepackage{comment}
\usepackage{amsthm}
\theoremstyle{plain}

\usepackage{lipsum}
\IEEEaftertitletext{\vspace{-2\baselineskip}}

\begin{document}
\bstctlcite{IEEEexample:BSTcontrol} 

\title{NOMA Joint Channel Estimation and Signal Detection using Rotational Invariant Codes and GMM-based Clustering
\vspace{-0.1em}}

%
%

\author{
{Ayoob~Salari,~\IEEEmembership{Student~Member,~IEEE}, Mahyar~Shirvanimoghaddam,~\IEEEmembership{Senior~Member,~IEEE},\\
Muhammad~Basit~Shahab,~\IEEEmembership{Member,~IEEE}, Yonghui~Li,~\IEEEmembership{Fellow ,~IEEE}, and Sarah~Johnson,~\IEEEmembership{Member,~IEEE} \vspace{-0.3em} }
\vspace{-0.3em} \thanks{A. Salari, M. Shirvanimoghaddam and Y. Li are with the School of Electrical and Information Engineering, The University of Sydney, Darlington, NSW 2006, Australia (e-mail: \{ayoob.salari, mahyar.shirvanimoghaddam,  yonghui.li\}@sydney.edu.au).}
\thanks{M. Shahab and S. Johnson are with the School of Engineering, University of Newcastle, Callaghan, NSW 2308, Australia (e-mail: \{basit.shahab, sarah.johnson\}@newcastle.edu.au).}
}

\maketitle


\begin{abstract}
This paper studies the joint channel estimation and signal detection for the uplink power-domain non-orthogonal multiple access. The proposed technique performs both detection and estimation without the need of pilot symbols by using a clustering technique. We apply rotational-invariant coding to assist signal detection at the receiver without sending pilot symbols. We utilize Gaussian mixture model (GMM) to automatically cluster the received signals without supervision and optimize decision boundaries to improve the bit error rate (BER) performance. Simulation results show that the proposed scheme without using any pilot symbol achieves almost the same BER performance as that for the conventional maximum likelihood receiver with full channel state information.

\end{abstract}


\IEEEpeerreviewmaketitle
\vspace{-2em}
\section{Introduction}
\vspace{-0.25em}
\IEEEPARstart{T}{he} fifth-generation (5G) of mobile communication aims to enable emerging applications with diverse range of requirements \cite{viswanathan2020communications}. In particular, massive machine-type communications (mMTC) and ultra-reliable low-latency communications (URLLC) have been considered as two other major use cases of 5G in addition to the enhanced mobile broadband (eMBB) \cite{5GNREricsson}. mMTC is expected to serve a massive number of low-complexity devices in a limited spectrum; hence, it should be robust, scalable, and energy efficient \cite{mahmood2020white}. 

As the orthogonal multiple access (OMA) is unable to meet the increasing demand for wireless access resulting from the exponential growth in the number of MTC devices, non-orthogonal multiple access (NOMA) has been proposed to allow users to share radio resources non-orthogonally, which enhances the spectral efficiency \cite{iswarya2021survey,shahab2020grant}. However, most of these approaches require channel state information (CSI) for power allocation and performing effective signal detection at the receiver \cite{ding2020unveiling}. 

Conventional communication systems use large block-length codes to approach the Shannon's capacity, assuming that the transmitter has access to CSI. Several channel estimation techniques have been proposed to effectively obtain CSI. These techniques are classified as blind, semi-blind, and training-based schemes \cite{nadeem2021non}. Using long pilot symbols, training-based channel estimation can estimate the channel accurately. On the other hand, at the cost of higher complexity, blind estimation takes advantage of the transmitted signal's features to estimate the channel in the absence of any training symbol. In many mMTC applications, each device usually sends a short packet of data; therefore, using long pilot signals will significantly reduce the system throughput. Moreover, estimating the channels to all active MTC devices is infeasible, therefore, devising novel grant-free NOMA approaches is favorable \cite{shahab2020grant}. Recently, the authors in \cite{salari2022clusteringbased,salari2020clustering} showed that without using long pilot sequences, the signal from paired users can be effectively detected using a clustering-based joint channel estimation and signal detection (JCESD) scheme \cite{salari2020clustering}. This approach still requires a few pilot symbols to determine the signal mapping in the constellation. In this letter, we take a step further to completely eliminate the need for any pilot symbol for the proposed JCESD scheme by using rotational-invariant (RI) codes. Our main contributions in this work are summarized next.  

1) Authors in \cite{salari2020clustering} demonstrated that by using the Gaussian mixture model (GMM) based clustering, we can cluster the received signal into $M$ clusters (for $M$ being the modulation order) in each stage of successive interference cancellation (SIC). By using the coordinates of cluster centroids, we can then estimate the channel gain $|h|$. However, this does not give us a clue about the phase of the channel. In \cite{salari2020clustering}, they sent a few pilot symbols to determine the demapping. Using such symbols will decrease the throughput, and should be avoided if possible.  In this paper, we show that by using rotational-invariant codes, we can easily eliminate the need for using any pilot symbol.

2) We also consider another channel coding technique, in which the low-density parity check (LDPC) code is used. To perform demapping without any pilot symbol, we consider all possible channel rotations, and calculate the syndrome for each channel phase. The phase which results in most checked syndromes is considered as the channel phase. Via simulations, we show that rotational-invariant codes outperform LDPC codes.

3) We further compare the proposed scheme with the maximum-likelihood receiver, in which a few pilot symbols are sent to estimate the channel. We showed that the ML receiver needs about 8 symbols to achieve almost the same BER performance as the proposed scheme without any pilot symbols. This means that under the same target BER, the ML receiver will loose about 16\% throughput, when each user packet size is 50 bits symbols (Assuming that we are using a rate-1/2 code with a 4-ary modulation).

The rest of this paper is structured as follows. Section II covers the system model and preliminaries. The proposed approach is detailed in the Section III. Simulation results are presented in Section IV followed by conclusion in Section VI.

\section{Background and System Model}
We study an uplink NOMA scenario, where $K$ users are paired to send their messages simultaneously to the BS.  User and the BS are equipped, each with single antenna. We assume that each user $i$ has a message $\mathbf{b}_i$ of length $k_u$ bits, which uses a rotationally invariant (RI) code $\mathcal{C}_{\mathrm{RI}}$ with rate $R$ to generate a codeword $\mathbf{c}_i$ of length $n_u=k_u/R$ bits. For the simplicity of notations, the encoding operation is denoted by $\mathbf{c}_i=\mathcal{C}_{\mathrm{RI}}(\mathbf{b}_i)$. Utilizing an $M$-ary modulation with signal constellation set $\mathcal{S}=\{\mathbf{s}_1,\cdots,\mathbf{s}_M\}$ where $\frac{1}{M}\sum_{i=1}^M|\mathbf{s}_i|^2=1$, each user $i$'s codeword will be modulated to generate a packet of length $N=n_u/\log_2(M)$ symbols, denoted by $\mathbf{x}_i$, and sent to the BS. The received superimposed signal at the BS, denoted by $\mathbf{y}\in \mathbb{C}^{N\times 1}$, can be expressed as:
\vspace{-0.5em}
\begin{equation} \label{eq:1}
    \mathbf{y} = \mathbf{X}\mathbf{h}  + \mathbf{w},
\end{equation}
where $\mathbf{X}=[\mathbf{x}_1,\cdots,\mathbf{x}_K] \in \mathcal{S}^{N\times K}$ denotes the transmitted signal matrix, $\mathbf{h}=[h_1,\cdots,h_K]\in \mathbb{C}^{K\times 1}$ is the channel vector, $\mathbf{w}\sim\mathcal{CN}(0, \nu\mathbf{I}_N)$ is the multivariate additive white Gaussian noise, $\mathbf{I}_N$ is the $N\times N$ identity matrix, and $\nu$ is the noise power. The channel between user $i$ and the BS, denoted by $h_i$, is modeled by a zero-mean circular symmetric complex Gaussian random variable, i.e., $h_i\sim\mathcal{CN}(0,1)$. We also consider block fading, that is the channel between each user and the BS remains unchanged for each transmission frame of length $N$ symbols. The signal-to-noise ratio (SNR) for user $u$, denoted by $\gamma_u$, is defined as $\gamma_u=|h_u|^2$. We further assume that the BS knows $N$, the total number of active users\footnote{This assumption can be relaxed as the proposed algorithm is iterative and can continue to detect all active users.}, and their modulation schemes, but the channel state information (CSI) is not available at users and BS, and BS attempts for joint channel estimation and signal detection.

\begin{figure}[t]
    \centering
    \includegraphics[width=0.8\columnwidth]{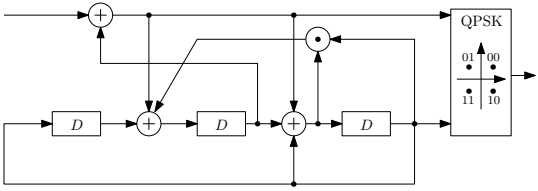}
    \caption{{$90$ degrees rate-$1/2$ RI  encoder}}
    \label{fig:Encoder}
\end{figure}%
\vspace{-1em}
\subsection{Rotational Invariant Code}
\vspace{-0.25em}
When the signal constellation has rotational symmetry, e.g. quadrature phase shift keying (QPSK), the presence of channel phase rotation combined with the absence of pilot symbols, results in the receiver having no knowledge of which constellation point has been sent \cite{pietrobon1994rotationally}. The time required to reestablish correct demodulation phase for decoding can cause channel outages during considerable
periods of time. In 1983, a full rotational invariant code by using nonlinear convolutional codes was proposed and applied to
trellis-coded modulation with a 32° point signal constellation \cite{ungerboeck1988codes}. In RI codes, all rotated versions of a modulated codeword are valid codeword sequences, and all the rotated versions of the same sequence are produced by the same information input and subsequently decode back to the same information. Fig. \ref{fig:Encoder} shows the encoder of a rate-$1/2$ $90$-degree RI code for QPSK modulation, where $D$ represents the unit delay element. This RI code is a non-linear convolutional code and can be effectively decoded by using the Viterbi decoder over the code trellis. Interested readers are referred to \cite{pietrobon1994rotationally} for further information on the design of RI codes. For notation simplicity, we denote the decoding process of the RI code by $\hat{\mathbf{b}}=\mathcal{C}^{-1}_{\mathrm{RI}}(\mathbf{c})$.

\vspace{-1em}
\section{The Proposed GMM-based Clustering algorithm for Joint Channel Estimation and Signal Detection Using RI Coding} 
\vspace{-0.25em}
Assuming that users' signals are randomly and uniformly drawn from the signal constellation $\mathcal{S}$, one can easily show that conditioned on the channel gains $\mathbf{h}$, the received signal at the BS at time instant $i$, denoted by $y_i$, have the following Gaussian mixture distribution \cite{salari2022clusteringbased}:
\begin{align} \label{eq:GMMmodel}
\vspace{-0.5em}
    y_i|\mathbf{h}\sim \frac{1}{|\mathcal{S}|^K}\sum_{\mathbf{s}\in{\mathcal{S}}^{K}}\mathcal{CN}\left(\mathbf{h}^T\mathbf{s},\nu\right),
\end{align}
where $|\mathcal{S}|$ is the modulation order. The Gaussian mixture model (GMM) \cite{singh2009statistical} is, therefore, a natural fit for our clustering problem. 
\begin{figure}[t]
\centering
\includegraphics[width=0.98\columnwidth]{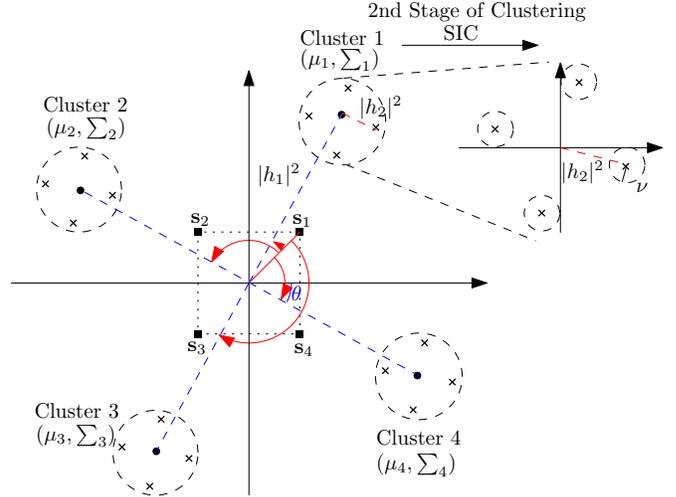}
\caption{The received signal at the BS in the I-Q plane for the 2-user NOMA scenario.}
\vspace{-2ex}
\label{fig:SchematicGMM}
\end{figure}

Fig. \ref{fig:SchematicGMM} shows the received signals at the BS in the I-Q plain for a 2-user NOMA scenario, when both users use the QPSK modulation. Without loss of generality, we assume that $|h_1|\ge |h_2|$. Due to fading, the received signal will be amplified and rotated with respect to the original constellation $\mathcal{S}=\{\mathbf{s}_1,\mathbf{s}_2,\mathbf{s}_3,\mathbf{s}_4\}$. The proposed JCESD approach, first cluster the received signals into $M=4$ clusters, where $M$ is the modulation order. By using the coordinates of cluster centroids,  the amplitude and phase of $h_1$ can be estimated. Then, the signals of user 1 will be detected. By using SIC, user 1's signal will be subtracted from the receive signal, and the algorithm again cluster the residual signal into $M=4$ clusters to estimate $h_2$ and detect user 2's signal. If more than 2 users are paired, the algorithm continues to detect all users' signals. 

It is important to note that estimating the channel phase in the above algorithm is not accurate. In particular, according to Fig. \ref{fig:SchematicGMM}, one can consider 4 different (and equally probable) choices for the phase of the channel, i.e., $\angle \hat{h}_1\in\{\theta-\pi/4, \theta-3\pi/4,\theta+\pi/4, \theta+3\pi/4\}$, where $-\pi/4\le\theta\le\pi/4$ is the phase of the cluster that resides in the angular region  $[-\pi/4,\pi/4]$. This means that we cannot demodulate signals to symbols. In [9], we considered that we send a few pilot symbols to determine the demapping. Using such symbols will decrease the throughput, and should be avoided if possible. This is the main issue that we are addressing in this letter, by using rotational invariant codes.

\vspace{-0.5em}
\subsection{Applying Clustering technique at the receiver}
\vspace{-0.25em}

When the channel gains are not known at the BS, we use an iterative clustering algorithm to cluster the received signal into $M$ clusters in each stage of SIC. Using the cluster's centroids, we can then estimate the channel gain and detect users' signals. We represent the probability density function of the two-dimensional multivariate normal distribution by $\mathcal{N}(\mathbf{z};\bm{\mu},\bm{\Sigma})$, where $\bm{\mu}$ is the vector of means and $\mathbf{\Sigma}$ denotes the covariance matrix. 
This method assumes that the data is created by a blend of Gaussian distributions. The mean, weight, and covariance of each Gaussian distribution component are parameterized by a GMM. When all users utilise the same $M$-ary modulation, there exist $M$ Gaussian distributions. Taking each distribution's weight as
$\omega_j$, $j\in\{1,\cdots,M\}$,  the mixed distribution may be expressed as a weighted sum of all Gaussian component densities, i.e.,
\begin{equation}\label{eq:3}
    \setlength\abovedisplayskip{0pt}
       p(\mathbf{z};\bm{\mu}_1,...,\bm{\mu}_M, \bm{\Sigma}_1,...\bm{\Sigma}_{M}) = \sum_{j=1}^{M} \omega_j \mathcal{N}_j(\mathbf{z};\bm{\mu}_j,\mathbf{\Sigma}_j)
       \setlength\belowdisplayskip{0pt},
\end{equation}
where $\sum_{j=1}^M\omega_j=1$. Given the observed data, we need to estimate the mean, weight and covariance of each distribution. We can accomplish this goal by maximizing the likelihood function over all available data at the BS \cite{salari2022clusteringbased}. To do this, we use the expectation maximisation (EM) technique \cite{hastie2009elements}, which is well-suited to addressing maximum likelihood problems involving hidden latent parameters.
Having a finite number of Gaussian mixtures, a closed-form expression for the parameters of the EM algorithm is possible \cite{dempster1977maximum}.

We start by initializing each Gaussian distribution's mean, weight, and covariance, and then we apply an iterative algorithm to calculate the parameters of the GMM. For each of the observations $i$ in the $t$-th iteration of algorithm, we assess the so-called responsibility variable of each model $j$ as follows:
\begin{align}\label{eq:4}
\setlength\abovedisplayskip{0pt}
        \hat{\gamma}_{i,j}^{(t)} = \frac{\hat{\omega_j}^{(t-1)} \mathcal{N}_j(\mathbf{z}_i;\hat{\bm{\mu}}_j^{(t-1)}, \hat{\mathbf{\Sigma}}_j^{(t-1)})}{\sum_{k=1}^{{M}} \hat{\omega_{k}}^{(t-1)} \mathcal{N}_{k}(\mathbf{z}_i;\hat{\bm{\mu}}_{k}^{(t-1)}, \hat{\mathbf{\Sigma}}_{k}^{(t-1)})}.
        \setlength\belowdisplayskip{0pt}
\end{align}
After that, we allocate each data point to its associated cluster. The computed responsibilities are then used to update the mean, weight, and covariance of each cluster in the next stage of the EM process as 
\begin{align}
       \setlength\abovedisplayskip{0pt}
       \label{eq:5}
       \hat{\omega}_j^{(t)}  &= \frac{ \sum_{i=1}^{N} {\hat{\gamma}}_{i,j}^{(t)}}{ \sum_{i=1}^{N} \sum_{k=1}^{{M}}  {\hat{\gamma}}_{i,k}^{(t)} },\\
       \label{eq:6}
       \hat{\bm{\mu}}_j^{(t)}  &= \frac{ \sum_{i=1}^{N} \hat{\gamma}_{i,j}^{(t)} \mathbf{z}_i }{ \sum_{i=1}^{N}  \hat{\gamma}_{i,j}^{(t)} },\\
       \label{eq:7}
       \hat{\mathbf{\Sigma}}_j^{(t)}  &= \frac{ \sum_{i=1}^{N} \hat{\gamma}_{i,j}^{(t)} (\mathbf{z}_i - \hat{\bm{\mu}}_j^{(t)}) (\mathbf{z}_i - \hat{\bm{\mu}}_j^{(t)})^T }{ \sum_{i=1}^{N}  \hat{\gamma}_{i,j}^{(t)} }. 
       \setlength\belowdisplayskip{0pt}
\end{align}
Finally, we evaluate the corresponding log-likelihood function and check for the convergence
\begin{align}\label{eq:8}
       \setlength\abovedisplayskip{0pt}
       &l^{(t)} (\bm{\mu}_1,\cdots,\bm{\mu}_{M}, \bm{\Sigma}_1,\cdots,\bm{\Sigma}_{M}|\mathbf{z}_1,\cdots,\mathbf{z}_N) \\ \nonumber
       & = \sum_{i=1}^{N} \ln \left[ \sum_{j=1}^{M}  \left( \omega_j^{(t)} \mathcal{N}_j(\mathbf{z}_i;\bm{\mu}_j^{(t)},\mathbf{\Sigma}_j^{(t)}) \right)  \right].
       \setlength\belowdisplayskip{0pt}
 \end{align}
It is assured that the EM method will converge to a local optima \cite{dempster1977maximum}. When users are using the same $M$-ary modulation and the users' signals are obtained evenly from the same constellation, the clusters' weight will be the same, i.e.,  $\omega_j = \frac{1}{M}$, $j=1,\cdots,M$. Furthermore, utilising QPSK modulation and a SIC receiver, we only need to estimate four Gaussian distributions at each step of the SIC. This contributes to the reduction of computational complexity.
\begin{algorithm}[t]
        \KwIn{Number of users $K$, received signal $\mathbf{y}$, modulation order $M$, Signal Constellation $\mathcal{S}$, and convergence threshold $\epsilon$}
        \KwOut{Decoded information of each user ($\hat{\mathbf{b}}_u, \forall u$ ).}

            Set $\omega_j = \frac{1}{M}, ~~j=1,\cdots,M$

         \For{ user $u= 1:K$ }
           {     
                \textbf{Initialize} Find $\hat{\bm{\mu}}^{(0)}$ and $\hat{\mathbf{\Sigma}}^{(0)}$ by clustering the received signal into $M$ clusters by sunning a few iterations of K-means clustering
                
                Compute $\hat{\gamma}_{i,j}^{(0)}$ (\ref{eq:4}), for $i=1:N$ and $j=1:M$
                
                Compute log-likelihoods based on (\ref{eq:8})
                
                Set $t=1$
               
            \While {$ l^{(t)} - l^{(t-1)} \geq \epsilon $}
                {
                Update $\hat{\bm{\mu}}^{(t)}$ and $\hat{\mathbf{\Sigma}}^{(t)}$ based on (\ref{eq:6}) and (\ref{eq:7})
                
                Update $\hat{\gamma}_{i,j}^{(t)}$ based on (\ref{eq:4})
                
                Update log-likelihoods based on (\ref{eq:8})
                }
            \textbf{Return} $\hat{\bm{\mu}}$=$\hat{\bm{\mu}}^{(t)}$ and $\hat{\mathbf{\Sigma}}=\hat{\mathbf{\Sigma}}^{(t)}$
            
            Find each cluster's phase: $\phi_i = \tan^{-1}\left(\frac{\mathrm{Im}(\hat{\mathbf{\mu}_i})}{\mathrm{Re}(\hat{\mathbf{\mu}_i})} \right)$
            
            Find clusters' average phase $\theta = \frac{1}{M}\sum_{i=1}^{M} {\phi_i} -\pi$ and channel amplitude $|\hat{h}_u|=\frac{1}{M}\sum_{i=1}^M \mathrm{abs}(\mu_i)$
            
            Update constellation decision boundaries based on $\theta$ and demodulate the signal to symbols: $\hat{\mathbf{x}}_u=demod(\mathbf{y})$
            
            Use the Viterbi algorithm to decode the user's information bit: $\hat{\mathbf{b}}_{u}=\mathcal{C}^{-1}_{\mathrm{RI}}(\hat{\mathbf{x}}_u)$.
            
            Re-encode this user, modulate and multiply by the estimated channel and subtract from superimposed received signal: 
            $\mathbf{y}\leftarrow\mathbf{y}-|\hat{h}_u|e^{j\theta} \mathcal{C}_{\mathrm{RI}}(\hat{\mathbf{b}}_u)$
    }
        \caption{GMM-based JCESD with RI Coding}
        \label{GMM Clustering Algorithm}
        \vspace{-0.25em}
\end{algorithm}

The proposed JCESD algorithm using RI coding is summarized in Algorithm \ref{GMM Clustering Algorithm}. First, we initialize the mean and covariance of GMM clustering by running a few iterations (less than five) of the K-means clustering algorithm \cite{hastie2009elements} (line 3). Then, we continue by finding the responsibility and log-likelihood function based on  (\ref{eq:4}) and (\ref{eq:8}), respectively (lines 4 and 5). After that, we attractively calculate the centroids of clusters and covariance matrices (lines 7 to 11 in the Algorithm \ref{GMM Clustering Algorithm}). 
After convergence, the phase of each cluster is obtained (line 13). To reduce the impact of phase rotation, we average the phase difference between the centroids of each cluster and their expected values (Step 14)\footnote{One can easily show that for a 2-user NOMA scenario, if the channels are known at the BS and $|h_1|\ge |h_2|$, for user 1, we have $\bm{\mu}^{\mathrm{opt}}_i=h_1\mathbf{s}_i$ and $\mathbf{\sum}^{\mathrm{opt}}_i=(|h_2|^2+\nu)\mathbf{I}_2$, and for user 2, we have $\bm{\mu}^{\mathrm{opt}}_i=h_2\mathbf{s}_i$, $\mathbf{\sum}^{\mathrm{opt}}_i=\nu\mathbf{I}_2$.}. We then adjust the decision boundaries and demodulate the received signal into symbols (line 15). We then use the Viterbi algorithm to decode the user's information (Step 16). Finally, we re-encode this user and modulate and multiply by the estimated channel and subtract it from the superimposed received signal (Step 17) and repeat the algorithm for the next user.

Algorithm \ref{GMM Clustering Algorithm} can be applied for the NOMA scenario with any number of users. Assuming the GMM converges after $t$ iterations, for the general case of $d$-dimensional $M$-ary modulations, the computational complexity
in the expectation step, which comprises of  calculating the determinant and the inverse of the covariance matrix, is in the order of $\mathcal{O}(MNd^3)$. The maximization step of GMM consists of computing the weight, mean, and covariance for each cluster with the associated complexity of $\mathcal{O}(MN)$, $\mathcal{O}(MN)$, and $\mathcal{O}(MNd)$, respectively. The overall complexity of the proposed GMM-based clustering will be in the order of $\mathcal{O}(tMd^3)$ operations per symbol \cite{salari2022clusteringbased}.
\vspace{-1.25em}
\section{Simulation Results}
\vspace{-0.25em}
\begin{figure*}[!t]
     \centering
     \begin{subfigure}[t]{0.32\textwidth}
         \centering
         \includegraphics[width=0.95\columnwidth]{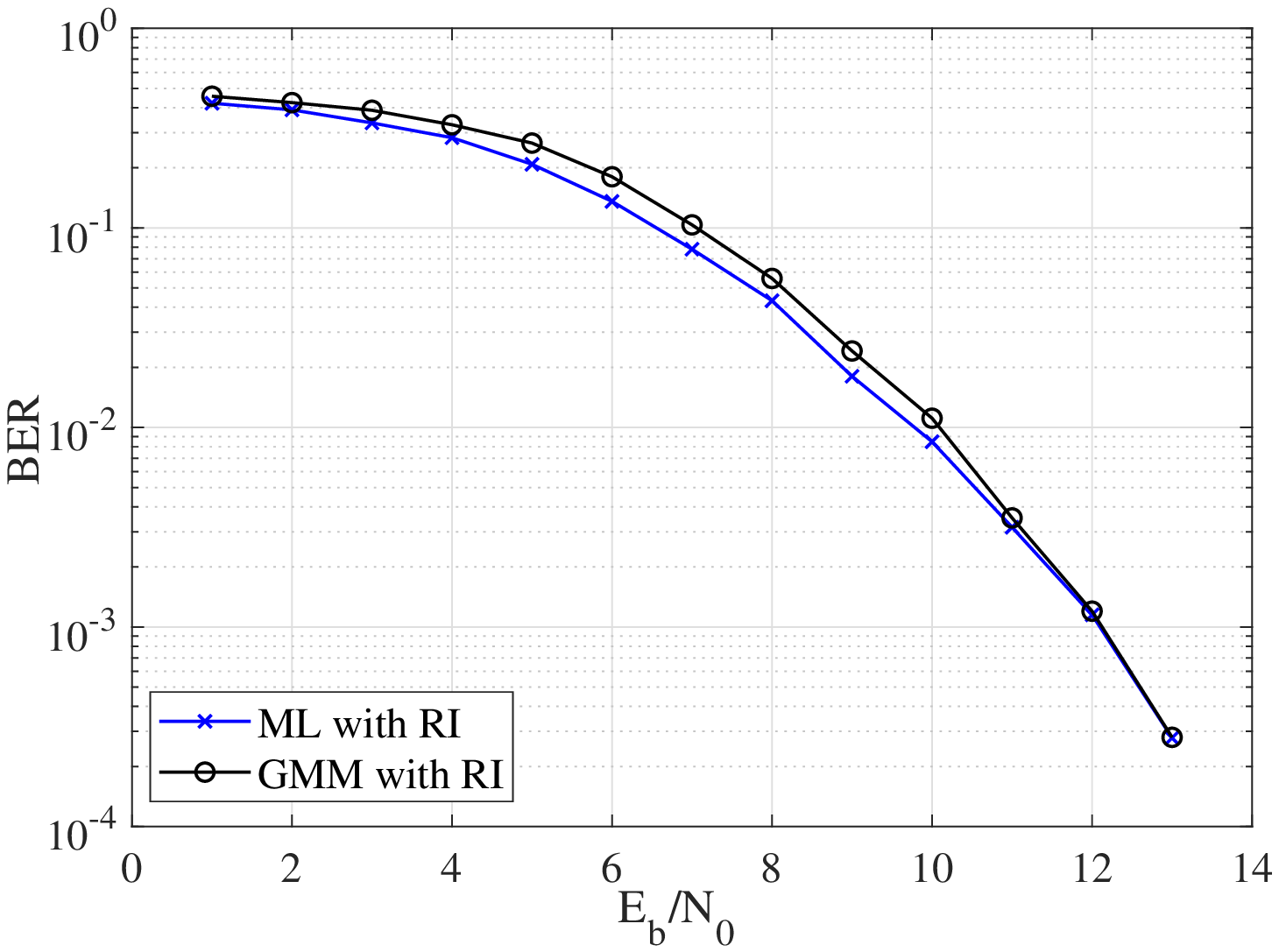}
         \caption{point-to-point}
         \label{fig:P2P}
     \end{subfigure}
     \hfill
     \begin{subfigure}[t]{0.32\textwidth}
         \centering
         \includegraphics[width=0.95\columnwidth]{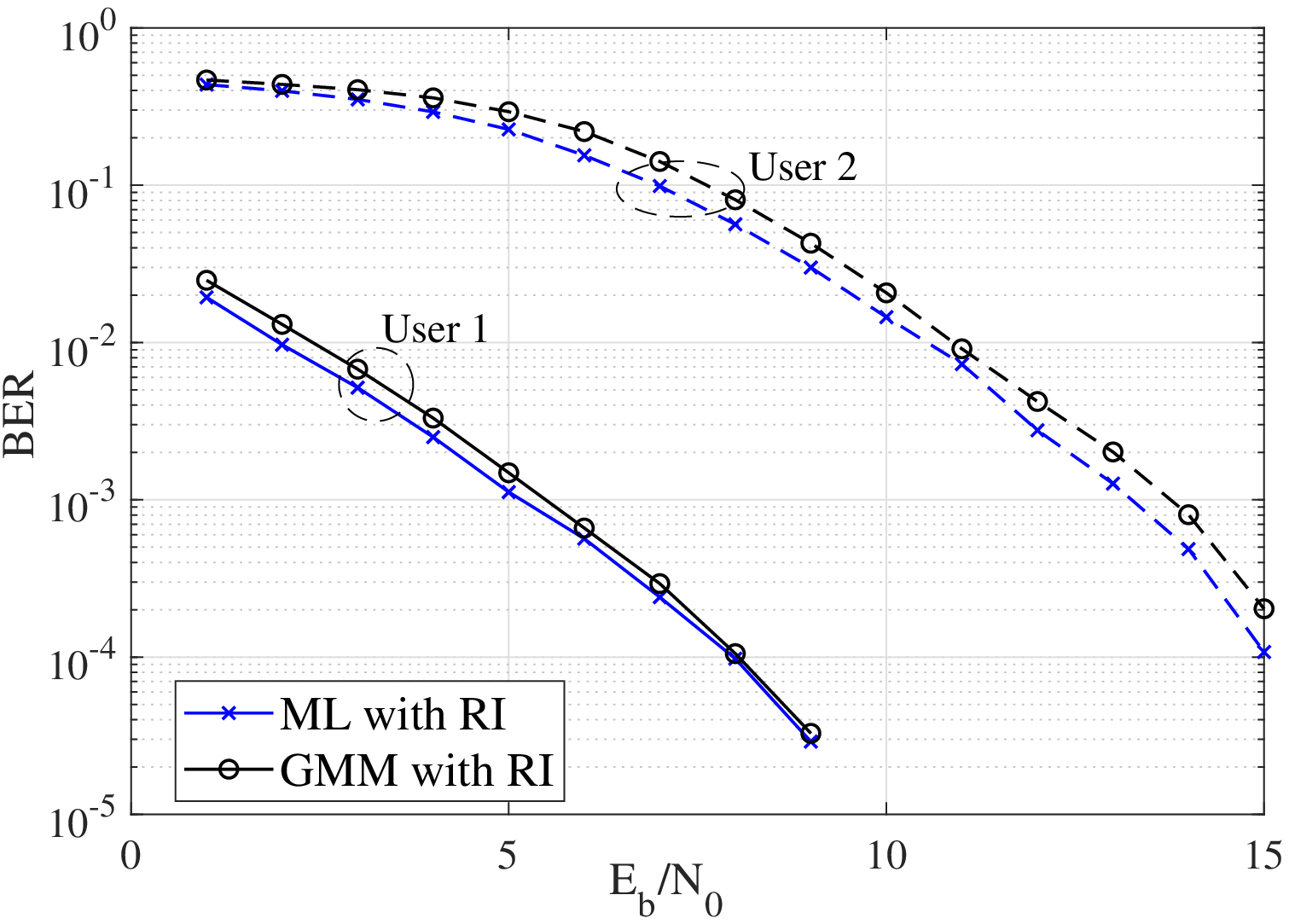}
         \caption{2-user NOMA, $\gamma_1-\gamma_2=9$dB.}
         \label{fig:NOMA}
     \end{subfigure}
     \hfill
     \begin{subfigure}[t]{0.32\textwidth}
         \centering
         \includegraphics[width=0.95\columnwidth]{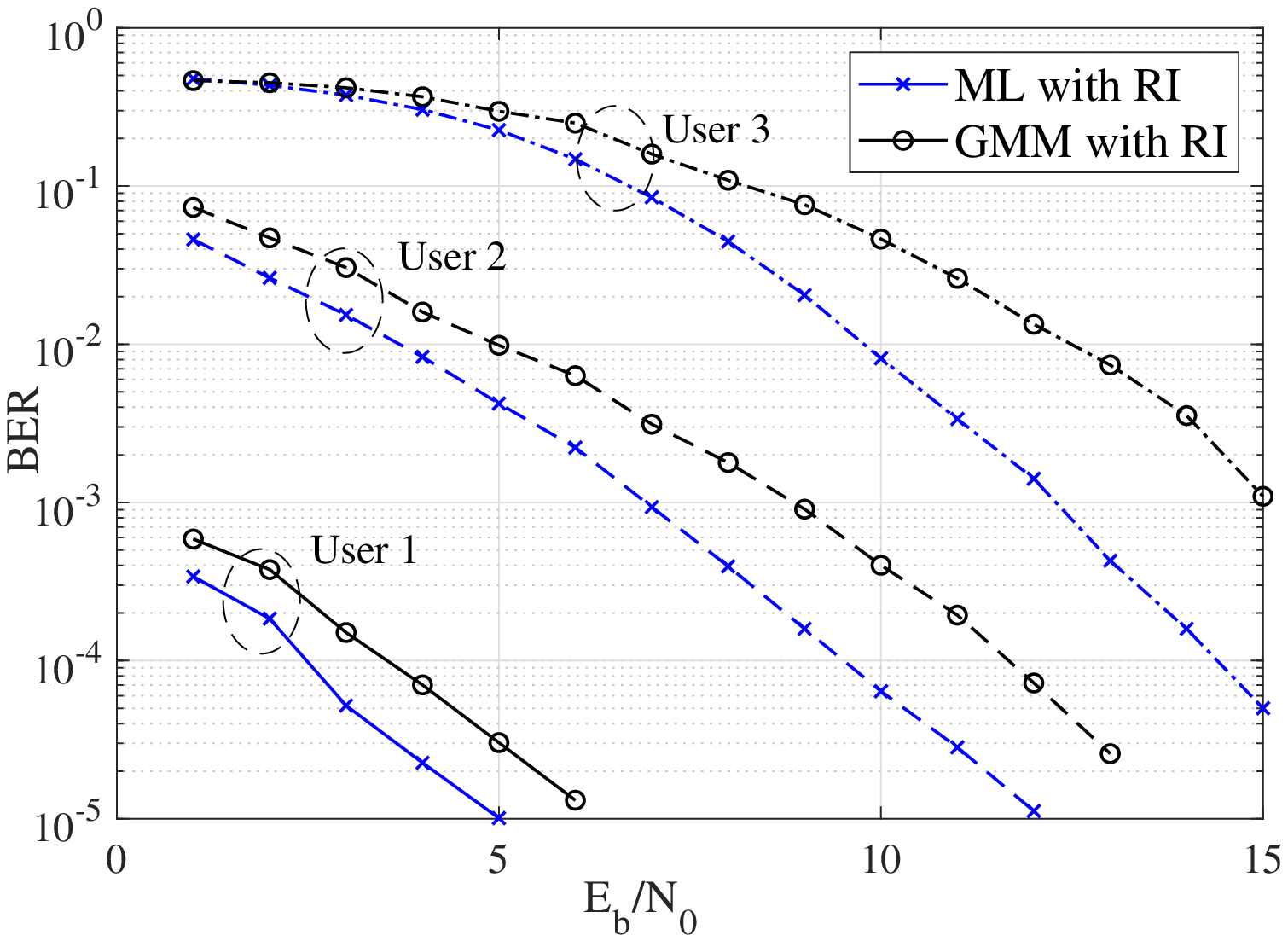}
         \caption{3-user NOMA, $\gamma_1-\gamma_2=9$dB, $\gamma_2-\gamma_3=9$dB.}
         \label{fig:3 User}
     \end{subfigure}
        \caption{{BER performance of the proposed algorithm vs. ML using rate 1/2 RI code (Fig. \ref{fig:Encoder}), when $N=50$.}}
        \label{fig:comp1}
\end{figure*}

In this section, we evaluate the performance of the proposed clustering-based JCESD algorithm utilizing RI codes. For all simulations, we assume that each user has a packet of length $k_u=50$ information bits to send to the BS, and a rate-1/2 RI code and QPSK modulation are employed by the users. 

\begin{figure}
    \centering
    \includegraphics[width=0.85\columnwidth]{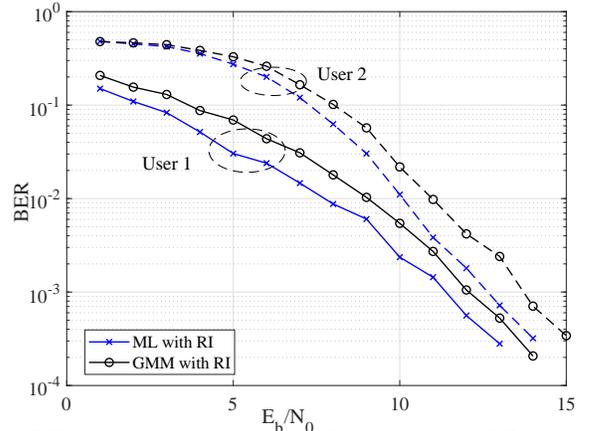}
    \vspace{-2ex}
    \caption{BER performance of the 2-user NOMA using rate 1/2 RI code, when $N=50$ and $\gamma_1-\gamma_2=7$ dB.}
    \label{fig:GMM7dB}
\end{figure}

Fig. \ref{fig:P2P} shows the bit error rate (BER) performance of the proposed JCESD approach using RI coding for a point-to-point communication scenario. As can be seen, the proposed approach performs very close to the ideal maximum-likelihood (ML) receiver with full CSI. Fig. \ref{fig:NOMA} shows the performance of the proposed approach in a 2-user NOMA scenario with a 9dB power difference between the users. As demonstrated in this figure, the proposed approach is capable of accurately determining clusters and performing symbol detection with a BER that is nearly equal to the performance of the optimal ML detection with full CSI at the BS.
 It is important to note that for all power-domain NOMA schemes, even with perfect ML detector and perfect CSI at the receiver, there should be a relatively high power differences between users, so that the receiver can successfully detect them. This limits the number of users that can be paired in NOMA scenarios to only 2 or 3 users. Fig. \ref{fig:3 User} shows the performance of the proposed approach to a 3-user NOMA scenario. The gap between the performance of the proposed approach and that for the optimal ML is mostly related to the sub-optimal decision boundaries found by GMM due to the limited observation. Using the QPSK modulation for three users, results in total 64 clusters. As we have considered packets of only 50 symbols, this means that there may exist some clusters without any signal point. In \cite{salari2022clusteringbased}, it has been demonstrated that when the sample size $N$ is sufficiently large, even when the number of users increases, the clustering technique performs extremely close to ML detection with full CSI. Fig. \ref{fig:GMM7dB} shows the performance of the proposed scheme when the gap between the received signal powers is 7 dB. While the gap to the ML detector with full CSI is slightly increased compared to that in Fig. \ref{fig:NOMA}, the proposed scheme can achieve a reasonable performance without the need for any pilot symbol. The gap can be reduced even if the power difference is lower, when $N$ is increased \cite{salari2022clusteringbased}.

\begin{figure}[t]
    \centering
    \includegraphics[width=0.8\columnwidth]{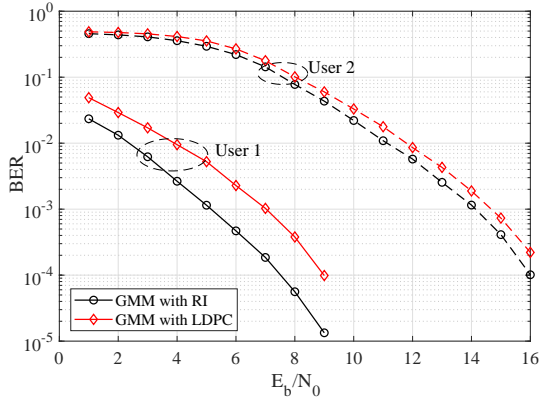}
    \vspace{-1ex}
    \caption{\small{BER comparison of RI code with rate 1/2 (Fig. \ref{fig:Encoder}) versus LDPC code with rate 1/2 using GMM. }}
    \label{fig:RI versus LDPC}
\end{figure}

For the sake of comparison, we consider another approach based on low-density parity check (LDPC) codes paired with the GMM-based clustering approach. The detection algorithm remains the same up until step 14 of Algorithm 1. We consider all possible (4 in case of QPSK) of channel rotations, and perform LDPC decoding and find the syndrome. The proper mapping may be determined by picking the mapping with the most syndrome checked. This method has a very high computational complexity, particularly when the number of users increases. Fig. \ref{fig:RI versus LDPC} shows the performance of the proposed JCESD algorithm with different coding techniques with rate $1/2$. As can be seen, JCESD with RI coding outperforms that with LDPC coding in terms of BER. For the strong user, the gap between the proposed RI coding technique and LDPC coding with GMM is more than 2dB at the BER of $10^{-3}$.

\begin{figure}[t]
    \centering
    \includegraphics[width=0.8\columnwidth]{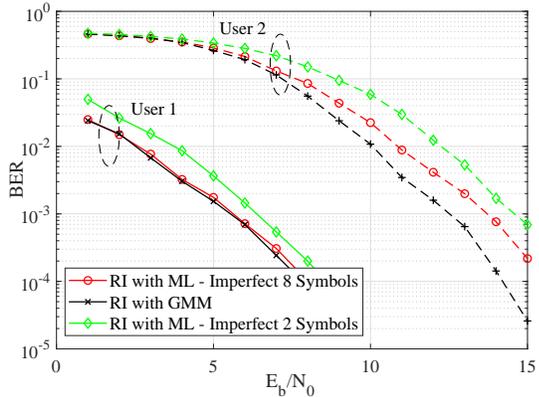}
    \vspace{-1ex}
    \caption{\small{BER comparison of the GMM-based scheme and imperfect-ML using rate 1/2 RI code (Fig. \ref{fig:Encoder}).}}
    \label{fig:Imperfect ML}
\end{figure}

For ML detection, attaining accurate CSI with a limited number of symbols is not viable in practice. As can be seen in Fig. \ref{fig:Imperfect ML}, the ML-based detection with only 2 training symbols performs poorly compared with our proposed technique without any pilot symbol. For ML to have a BER performance similar to our proposed scheme, at least 8 training symbols should be used. However, this results in about 16\% loss in throughput. Moreover, even with 8 training symbols, our proposed approach has better performance for the weak user.

\begin{figure}[t]
    \centering
    \includegraphics[width=0.8\columnwidth]{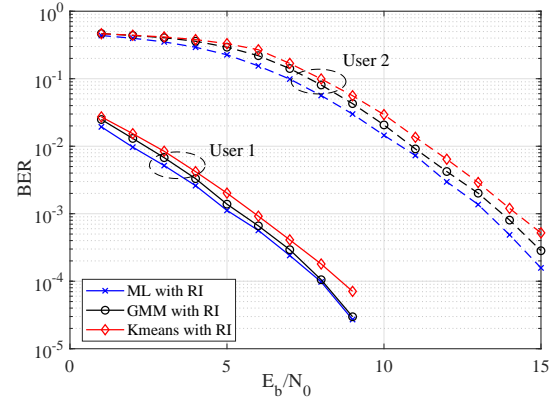}
    \vspace{-1ex}
    \caption{\small{BER comparison of ML vs GMM vs Kmeans for Two user NOMA using rate 1/2 RI code (Fig. \ref{fig:Encoder}).}}
    \label{fig:ML vs GMM vs Kmeans}
\end{figure}

Fig. \ref{fig:ML vs GMM vs Kmeans} shows the BER performance of 2-user NOMA with RI coding for GMM clustering versus K-means clustering. As opposed to GMM clustering in Algorithm 1, K-means assumes that the covariance matrix for all clusters is the same. As can be seen, the GMM-based clustering outperforms the K-mean clustering approach.
\vspace{-0.75em}
\section{Conclusion}
\vspace{-0.25em}
By combining clustering technique with rotational-invariant coding, this paper investigated the JCESD for the uplink of non-orthogonal multiple access without the use of any pilot symbols for channel estimation. In order to counteract the effects of channel rotation, we employ rotational invariant coding, which allows us to communicate without the use of pilot signals. We employ the Gaussian mixture model to cluster incoming signals without supervision, and optimise decision boundaries in accordance with the clustering results in order to improve the bit error rate (BER). Results showed that the proposed scheme without any pilot symbols can achieve the same performance as the maximum-likelihood detector that needs to obtain full CSI to operate well.
\vspace{-0.75em}
\bibliographystyle{IEEEtran}
\bibliography{IEEEabrv,BIB}

\begin{thebibliography}{10}
\providecommand{\url}[1]{#1}
\csname url@samestyle\endcsname
\providecommand{\newblock}{\relax}
\providecommand{\bibinfo}[2]{#2}
\providecommand{\BIBentrySTDinterwordspacing}{\spaceskip=0pt\relax}
\providecommand{\BIBentryALTinterwordstretchfactor}{4}
\providecommand{\BIBentryALTinterwordspacing}{\spaceskip=\fontdimen2\font plus
\BIBentryALTinterwordstretchfactor\fontdimen3\font minus
  \fontdimen4\font\relax}
\providecommand{\BIBforeignlanguage}[2]{{%
\expandafter\ifx\csname l@#1\endcsname\relax
\typeout{** WARNING: IEEEtran.bst: No hyphenation pattern has been}%
\typeout{** loaded for the language `#1'. Using the pattern for}%
\typeout{** the default language instead.}%
\else
\language=\csname l@#1\endcsname
\fi
#2}}
\providecommand{\BIBdecl}{\relax}
\BIBdecl

\bibitem{viswanathan2020communications}
H.~Viswanathan and P.~E. Mogensen, ``White paper: Communications in the 6{G}
  {E}ra,'' \emph{Nokia Bell Labs}, 2020.

\bibitem{5GNREricsson}
J.~Peisa, P.~Persson, S.~Parkvall, E.~Dahlman, A.~Grovlen, C.~Hoymann, and
  D.~Gerstenberger, ``{5G New Radio Evolution},'' \emph{Ericsson Technology
  Review}, no.~2, March 2020.

\bibitem{mahmood2020white}
N.~H. Mahmood, S.~B{\"o}cker \emph{et~al.}, ``White paper on critical and
  massive machine type communication towards 6{G},'' \emph{arXiv:2004.14146},
  2020.

\bibitem{iswarya2021survey}
N.~Iswarya and L.~Jayashree, ``A survey on successive interference cancellation
  schemes in non-orthogonal multiple access for future radio access,''
  \emph{Wireless Personal Communications}, pp. 1--22, 2021.

\bibitem{shahab2020grant}
M.~B. Shahab, R.~Abbas, M.~Shirvanimoghaddam, and S.~J. Johnson, ``Grant-free
  non-orthogonal multiple access for {IoT}: A survey,'' \emph{IEEE
  Communications Surveys \& Tutorials}, vol.~22, pp. 1805--1838, 2020.

\bibitem{ding2020unveiling}
Z.~Ding, R.~Schober, and H.~V. Poor, ``Unveiling the importance of {SIC} in
  {NOMA} systems—part 1: State of the art and recent findings,'' \emph{IEEE
  Communications Letters}, vol.~24, no.~11, pp. 2373--2377, 2020.

\bibitem{nadeem2021non}
F.~Nadeem, M.~Shirvanimoghaddam, Y.~Li, and B.~Vucetic, ``Non-orthogonal {HARQ}
  for {URLLC}: Design and analysis,'' \emph{IEEE Internet of Things Journal},
  2021.

\bibitem{salari2022clusteringbased}
A.~Salari, M.~Shirvanimoghaddam, M.~B. Shahab, R.~Arablouei, and S.~Johnson,
  ``Clustering-based joint channel estimation and signal detection for
  {NOMA},'' \emph{arXiv preprint arXiv:2201.06245}, 2022.

\bibitem{salari2020clustering}
A.~Salari, M.~Shirvanimoghaddam, M.~B. Shahab, R.~Arablouei, and S.~Johnson,
  ``Clustering-based joint channel estimation and signal detection for
  grant-free {NOMA},'' in \emph{IEEE Globecom Workshops}, 2020, pp. 1--6.

\bibitem{pietrobon1994rotationally}
S.~S. Pietrobon, G.~Ungerboeck, L.~C. P{\'e}rez, and D.~Costello,
  ``Rotationally invariant nonlinear trellis codes for two-dimensional
  modulation,'' \emph{IEEE Transactions on Information Theory}, vol.~40, no.~6,
  pp. 1773--1791, 1994.

\bibitem{ungerboeck1988codes}
G.~Ungerboeck and S.~S. Pietrobon, ``Codes for {QPSK} modulation with
  invariance under 90 degrees rotation,'' in \emph{Jet Propulsion Lab.,
  Proceedings of the Mobile Satellite Conference}, 1988.

\bibitem{singh2009statistical}
R.~Singh, B.~C. Pal, and R.~A. Jabr, ``Statistical representation of
  distribution system loads using {G}aussian mixture model,'' \emph{IEEE
  Transactions on Power Systems}, vol.~25, no.~1, pp. 29--37, 2009.

\bibitem{hastie2009elements}
T.~Hastie, R.~Tibshirani, and J.~Friedman, \emph{The elements of statistical
  learning: data mining, inference, and prediction}.\hskip 1em plus 0.5em minus
  0.4em\relax Springer Science \& Business Media, 2009.

\bibitem{dempster1977maximum}
A.~P. Dempster, N.~M. Laird, and D.~B. Rubin, ``Maximum likelihood from
  incomplete data via the {EM} algorithm,'' \emph{Journal of the Royal
  Statistical Society: Series B}, vol.~39, no.~1, pp. 1--22, 1977.

\end{thebibliography}

\end{document}